\documentclass[sigconf]{acmart}

\usepackage{adjustbox}
\usepackage{booktabs}
\usepackage{multirow}
\usepackage{xcolor}
\usepackage{tikz}
\usetikzlibrary{positioning}

\AtBeginDocument{%
  }

\copyrightyear{2026}
\acmYear{2026}
\acmDOI{XXXXXXX.XXXXXXX}

\begin{document}

\title{Corporate Training in Brazilian Software Engineering: A Quantitative Study of Professional Perceptions}

\author{Rodrigo Siqueira}
\orcid{0009-0004-6755-9746}
\affiliation{%
  \institution{CESAR School}
  \city{Recife}
  \state{PE}
  \country{Brazil}
}
\email{rms9@cesar.school}

\author{Antonio Oliveira}
\affiliation{
  \institution{CESAR School}
  \city{Recife}
  \state{PE}
  \country{Brazil}
}
\email{aaspo@cesar.school}

\author{Breno Alves de Andrade}
\affiliation{
  \institution{Cesar School}
  \city{Recife}
  \state{PE}
  \country{Brazil}
}
\email{baa3@cesar.school}

\author{Lidiane C S Gomes}
\affiliation{
  \institution{CESAR School}
  \state{PE}
  \city{Recife}
  \country{Brazil}
}
\email{lcsg@cesar.school}

\author{Danilo Monteiro Ribeiro}
\affiliation{
  \institution{CESAR School}
  \city{Recife}
  \state{PE}
  \orcid{0000-0001-7393-729X}
  \country{Brazil}
}
\email{dmr@cesar.school}

\renewcommand{\shortauthors}{Siqueira et al.}

\begin{abstract}
\textbf{Context:} Strategic corporate training is essential for the sustained professional development of software engineers. However, there is a knowledge gap regarding the factors that drive quality and effectiveness of such training from the professionals' perspective, and no validated instrument exists for assessing these factors in the software engineering (SE) domain.
\textbf{Objective:} This study aims to quantitatively analyze which factors influence SE professionals' perceptions of corporate training quality and effectiveness.
\textbf{Method:} A quantitative survey was conducted with 282 Brazilian SE professionals. A structured questionnaire was developed and polychoric correlation was adopted for data analysis.
\textbf{Results:} Three tightly correlated factors (cognitive engagement, variety of activities, and instructor performance) emerged as the strongest predictors of perceived training quality and effectiveness. Mandatory participation significantly reduces motivation and perceived training quality. Perceived impact on personal time proved to be largely independent of training quality. These findings are consistent with the general training effectiveness literature.
\textbf{Conclusions:} Training effectiveness in the SE context is predominantly determined by three factors: cognitive engagement, variety of activities, and instructor performance. Mandatory participation negatively influences motivation, perceived relevance, and perceived training quality, while also amplifying the perception of time burden. The consistency with the general literature suggests that software organizations do not need to reinvent training design principles and can apply established guidelines with confidence. Salas and Cannon-Bowers' framework produced coherent results in the SE context, making it a promising candidate for future psychometric validation.
\end{abstract}




\maketitle

\section{Introduction}\label{section:introduction}

Corporate training that is perceived as relevant and aligned with real work demands has a strong potential to improve both individual and organizational performance \cite{coverstone2003training}. Training effectiveness depends not only on content or format, but on a complex interaction of individual, instructional, and organizational factors \cite{salas2001science}. Previous research has shown that learning motivation, manager support, and organizational climate are central to supporting learning and knowledge transfer \cite{facteau1995influence}.

Despite growing training investments in the software sector, important gaps remain in understanding how professionals perceive the quality and effectiveness of these programs \cite{de2025mapping}. The Human Resources literature has given limited empirical attention to the complexity of training transfer, focusing more on organizational policies than on the participant's experience \cite{santos2003employee}. Evidence from the software industry professional's perspective is especially scarce \cite{de2025mapping}.

Software engineering has distinctive characteristics that make corporate training not just an educational benefit, but a core concern for engineering practice. Technologies evolve in cycles of months rather than decades, making continuous professional development a prerequisite for productive teams and competitive products \cite{diniz2024skill}. The well-documented gap between academic curricula and industry demands places an added burden on corporate training to bridge skill gaps \cite{diniz2024skill}. The quality of such training therefore has direct consequences for software quality, team productivity, and defect rates \cite{devaraj2004measure, assyne2022state}, outcomes that are central to software engineering research, not only to educational scholarship.

Of special interest is understanding how the nature of training (mandatory or voluntary) affects engagement and motivation \cite{gegenfurtner2016voluntary}, and which factors best predict perceived training quality and effectiveness. Answering these questions provides concrete guidance for software organizations designing or improving their training programs.

The central objective of this study is to quantitatively analyze how software engineering professionals working in Brazil perceive the quality and effectiveness of corporate training. This objective is addressed through one main research question and two sub-questions:

\begin{quote}
    \textbf{RQ1 —} Which factors determine software engineering professionals' perceptions of corporate training quality and effectiveness?
\end{quote}

\begin{quote}
    \textbf{RQ1.1 —} How does the nature of participation (mandatory versus voluntary) influence these perceptions?
\end{quote}

\begin{quote}
    \textbf{RQ1.2 —} Are these factors consistent with those reported in the general training literature?
\end{quote}

The systematic mapping by \citet{de2025mapping} identified the absence of validated instruments for assessing training effectiveness from the SE professional's perspective, making this operationalization a necessary step toward filling that gap.

This study offers the following main contributions:

\begin{enumerate}
    \item Identifies a core of three strongly correlated factors (problem-solving reasoning, variety of activities, and instructor performance) as the central predictors of perceived training quality and effectiveness among software professionals, providing organizations with a concrete prioritization framework when resources are limited.
    \item Provides quantitative empirical evidence that mandatory participation reduces motivation and perceived training quality, while voluntary participation favors engagement, consistent with findings in other domains.
    \item Compares these findings with the general training literature, and the observed consistency suggests that established training design principles can be applied to the software industry with confidence.
    \item Operationalizes Salas and Cannon-Bowers' training framework in a software industrial environment through a large-scale survey ($n=282$). The resulting instrument is exploratory and is offered as an open candidate for future psychometric validation and refinement.
\end{enumerate}

The remainder of this article is organized as follows: Section~2 presents the theoretical framework and related work; Section~3 describes the methodology; Section~4 presents the quantitative results; Section~5 discusses the main findings; Section~6 addresses limitations; and Section~7 concludes the article.

\section{Background}\label{section:background}

\subsection{Salas' Framework}\label{subsection:salas-framework}

We adopted the framework by \citet{salas2001science} as an analytical lens for this study. The framework offers a comprehensive view of organizational training, consolidating advances from the 1990s and 2000s and revised in 2012 \cite{salas2012science}. It emphasizes that training effectiveness results from the interaction of cognitive, motivational, and organizational factors, combined with appropriate instructional methods and post-training support \cite{salas2001science}. We operationalized its four main dimensions into questionnaire blocks, capturing perceptions related to: (i)~Training Needs Analysis; (ii)~Antecedent Conditions; (iii)~Training Methods/Strategies; and (iv)~Post-Training Conditions. This structure guided the construction of the survey instrument, the analysis of items, and the interpretation of findings. Table~\ref{tab:salas_categories} presents the four dimensions alongside the distribution of previously identified studies per sub-area, as reported in \citet{de2025mapping}.

Several alternative frameworks were considered. Kirkpatrick's four-level model \cite{kirkpatrick1970evaluation} is widely used for training evaluation, organizing assessment into reaction, learning, behavior, and results. However, it is primarily an \textit{evaluation} framework that defines \textit{what to measure}, rather than identifying \textit{which factors influence} training effectiveness, which is the focus of this study's research question. Baldwin and Ford's transfer model \cite{baldwin1988transfer} addresses an important part of the training cycle (trainee characteristics, training design, and work environment as predictors of transfer), but it focuses specifically on post-training transfer rather than the full training lifecycle. Salas and Cannon-Bowers' framework was chosen because it covers the entire training process, from needs analysis through post-training conditions, providing the broadest analytical structure for exploring which factors across all dimensions shape SE professionals' perceptions.

\begin{table}[ht]
    \centering
    \caption{Categories, Subcategories, and Distribution of Studies in Salas' Framework}
    \label{tab:salas_categories}
    \small
    \renewcommand{\arraystretch}{1.4}
    \begin{tabular}{| p{0.26\linewidth} | p{0.50\linewidth} | c |}
        \hline
        \textbf{Category} & \textbf{Subcategories} & \textbf{Qty.} \\
        \hline
        \multirow{2}{=}{Training Needs Analysis}
            & \textbf{Organizational Analysis}: Focuses on alignment with strategic objectives, resources, and constraints.
            & 3 \\
            \cline{2-3}
            & \textbf{Job/Task Analysis}: Identifies specific Knowledge, Skills, and Attitudes (KSAs) required for effective task performance.
            & 0 \\
        \hline
        \multirow{3}{=}{Antecedent Training Conditions}
            & \textbf{Individual Characteristics}: Traits such as cognitive ability, self-efficacy, and goal orientation.
            & 2 \\
            \cline{2-3}
            & \textbf{Training Motivation}: Shaped by individual and organizational variables; drives retention and behavioral change.
            & 1 \\
            \cline{2-3}
            & \textbf{Training Induction and Pretraining Environment}: Preparation strategies to optimize readiness and learning conditions.
            & 3 \\
        \hline
        \multirow{4}{=}{Training Methods and Instructional Strategies}
            & \textbf{Specific Learning Approaches}: Pedagogical techniques like feedback loops and reinforcement.
            & 6 \\
            \cline{2-3}
            & \textbf{Learning Technologies and Distance Training}: Use of digital tools (e-learning, video conferencing) for flexible delivery.
            & 4 \\
            \cline{2-3}
            & \textbf{Simulation-Based Training and Games}: Immersive experiences to reduce errors and improve performance.
            & 0 \\
            \cline{2-3}
            & \textbf{Team Training}: Focuses on collaborative skills and group effectiveness.
            & 2 \\
        \hline
        \multirow{2}{=}{Post-Training}
            & \textbf{Training Evaluation}: Measurement of effectiveness through behavioral, cognitive, and affective indicators.
            & 4 \\
            \cline{2-3}
            & \textbf{Transfer of Training}: Application, generalization, and maintenance of KSAs in the workplace.
            & 1 \\
        \hline
        \multicolumn{2}{|r|}{\textbf{Total}} & \textbf{26} \\
        \hline
    \end{tabular}
\end{table}

\subsection{Previous Systematic Mapping}\label{subsection:mapping}

A previous systematic mapping by \citet{de2025mapping} analyzed 26 primary studies and found that research focuses mainly on teaching methods and strategies, with significant gaps in other training dimensions. Table~\ref{tab:salas_categories} shows the distribution of studies by category. Notably, no studies were found in the subcategories \textit{Job/Task Analysis} and \textit{Simulation-Based Training and Games}, and few addressed \textit{Transfer of Training}.

\subsection{Related Work}\label{subsection:related_work}

The literature presents contrasting views on mandatory versus voluntary training participation. \citet{gegenfurtner2016voluntary} argue that voluntariness favors autonomous motivation and learning transfer, especially among participants with a learning orientation. In contrast, \citet{baldwin1991perils} suggest that mandatoriness often signals the strategic importance of the content, while too much voluntariness can be read as low organizational priority. More recently, \citet{de2025effects} showed, in a study with 1,122 trainees, that the effects of mandatory versus voluntary participation on transfer depend on whether training covers soft- or hard-skills, while voluntary participation is consistently better for transfer motivation.

Since the mandatory--voluntary distinction is central to this study's research questions, it is important to operationalize both concepts. Drawing on the literature, we adopt the following working definitions:

\textbf{Mandatory training} is a formal organizational requirement driven by legal obligations, internal regulations, or operational needs \cite{matulcikova2022further, facteau1995influence}. Its purpose is to ensure uniform skill development across a group or the entire organization. As \citet{facteau1995influence} show, compliance-driven attendance tends to reduce pretraining motivation; mandating training may get employees to attend but lower their motivation to learn. However, \citet{baldwin1991perils} and \citet{tsai2003perceived} note that mandatoriness can also signal that the organization considers the content strategically important.

\textbf{Voluntary training} is a development opportunity based on individual choice and self-motivation \cite{curado2015voluntary, gegenfurtner2016voluntary}. The topics offered typically reflect the organization's long-term strategic goals, allowing employees to grow in directions the organization values. Companies commonly fund materials and tuition but do not always compensate the employee's time \cite{curado2015voluntary}. \citet{curado2015voluntary} found that employees who enrolled voluntarily showed significantly higher autonomous motivation to transfer than those enrolled mandatorily, consistent with self-determination theory \cite{gegenfurtner2016voluntary}. \citet{de2025effects} further show that voluntary participation is especially beneficial for soft-skill trainings, where trainees experience greater autonomy and higher transfer.

Training effectiveness is multidimensional. \citet{fawad2012integrated} propose an integrated model with four dimensions: session satisfaction, content relevance, instructor performance, and learning transfer. In the software sector, \citet{devaraj2004measure} reinforce that technical content quality and direct applicability to job performance are key for training to be perceived as valuable. The instructor role has received particular attention: \citet{yaqoot2021predicted} show that trainer competence and training environment quality are significant predictors of effectiveness, especially in face-to-face and technical settings.

A theoretical foundation for understanding why certain factors influence training effectiveness more than others can be found in \citet{bandura1977social}'s Social Learning Theory. Bandura's observational learning model identifies four interdependent subprocesses that determine whether observed behavior is successfully acquired and reproduced: (1)~\textit{attention}, governed by the quality and attractiveness of the model; (2)~\textit{retention}, the ability to encode and store observed behavior symbolically; (3)~\textit{motor reproduction}, the capacity to convert symbolic representations into action; and (4)~\textit{motivation}, sustained by reinforcement and engagement with varied stimuli. In corporate training contexts, these subprocesses map naturally onto key instructional factors: the instructor serves as the model that captures attention, cognitive engagement supports retention through symbolic encoding, practical application enables reproduction, and varied activities sustain motivation throughout the learning process.

The gap between academic education and market demands is identified by \citet{diniz2024skill} as a main cause of the skill gap in the software industry. Closing these gaps requires training evaluation that goes beyond immediate reactions and focuses on direct impact on job performance \cite{devaraj2004measure}. Furthermore, \citet{facteau1995influence} show that social support from supervisors and peers has a stronger positive influence on transfer than extrinsic incentives, a finding supported by \citet{santos2003employee}.

Taken together, these studies show that corporate training in software engineering is not just a general educational question. The fast obsolescence of technical knowledge \cite{assyne2022state}, the structural gap between academia and industry \cite{diniz2024skill}, and the growing importance of continuous skills development in software engineering \cite{gegenfurtner2016voluntary, borges2024skills} create a training ecosystem that differs from those studied in traditional organizational psychology or generic HR research. A similar cross-domain movement has occurred in healthcare, where \citet{salas2009critical} identified that the same training effectiveness factors (leadership support, instructor qualification, learning climate, and trainee motivation) proved critical when applying organizational training principles to medical teams. Understanding how these dynamics apply to SE professionals is therefore a contribution to software engineering research, not only to educational scholarship.

\section{Method}\label{section:method}

\subsection{Study Design}

An empirical quantitative study was conducted using a cross-sectional survey design. Data were collected through a single structured online questionnaire and analyzed using polychoric correlation to explore associations among ordinal scale items and to identify predictors of perceived training quality and effectiveness.

\subsection{Instrument Design and Operationalization}

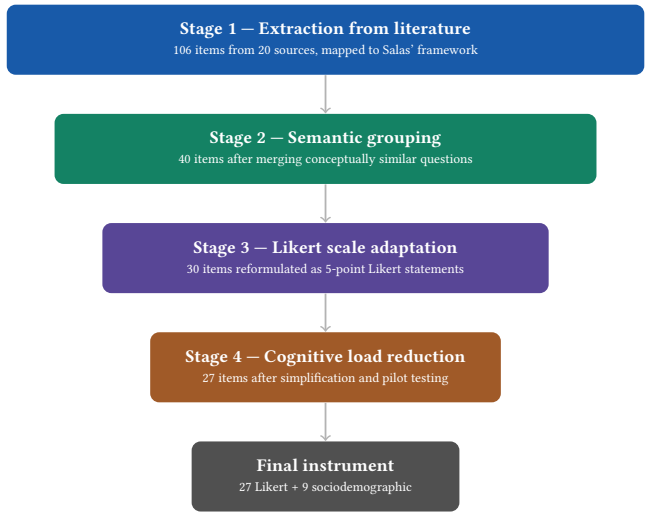
\begin{figure}[ht]
    \centering
    \resizebox{\columnwidth}{!}{%
    \begin{tikzpicture}[
        stage/.style={rectangle, rounded corners=4pt, minimum height=1.1cm, text centered, font=\small, text=white},
        arrow/.style={->, thick, gray!60},
    ]
    % Stage 1
    \node[stage, fill={rgb,255:red,24;green,90;blue,169}, minimum width=10cm] (s1) {
        \begin{tabular}{c}
            \textbf{Stage 1 --- Extraction from literature}\\[-1pt]
            {\scriptsize 106 items from 20 sources, mapped to Salas' framework}
        \end{tabular}
    };
    % Stage 2
    \node[stage, fill={rgb,255:red,20;green,130;blue,100}, minimum width=8.5cm, below=0.6cm of s1] (s2) {
        \begin{tabular}{c}
            \textbf{Stage 2 --- Semantic grouping}\\[-1pt]
            {\scriptsize 40 items after merging conceptually similar questions}
        \end{tabular}
    };
    % Stage 3
    \node[stage, fill={rgb,255:red,88;green,70;blue,150}, minimum width=7cm, below=0.6cm of s2] (s3) {
        \begin{tabular}{c}
            \textbf{Stage 3 --- Likert scale adaptation}\\[-1pt]
            {\scriptsize 30 items reformulated as 5-point Likert statements}
        \end{tabular}
    };
    % Stage 4
    \node[stage, fill={rgb,255:red,160;green,90;blue,40}, minimum width=5.5cm, below=0.6cm of s3] (s4) {
        \begin{tabular}{c}
            \textbf{Stage 4 --- Cognitive load reduction}\\[-1pt]
            {\scriptsize 27 items after simplification and pilot testing}
        \end{tabular}
    };
    % Final
    \node[stage, fill={rgb,255:red,80;green,80;blue,80}, minimum width=4.2cm, below=0.6cm of s4] (final) {
        \begin{tabular}{c}
            \textbf{Final instrument}\\[-1pt]
            {\scriptsize 27 Likert + 9 sociodemographic}
        \end{tabular}
    };
    % Arrows
    \draw[arrow] (s1) -- (s2);
    \draw[arrow] (s2) -- (s3);
    \draw[arrow] (s3) -- (s4);
    \draw[arrow] (s4) -- (final);
    \end{tikzpicture}%
    }%
    \caption{Overview of the four-stage instrument development process.}
    \label{fig:instrument_stages}
\end{figure}

The instrument was developed through a four-stage iterative process, guided by the dimensions of Salas and Cannon-Bowers' framework \cite{salas2001science,salas2012science}. The complete question matrix documenting all stages is available in the replication package (Section~\ref{sec:artefact}). Figure~\ref{fig:instrument_stages} provides an overview of the process.

\textbf{Stage~1 — Extraction from literature (106 items).} A search was conducted on Google Scholar using terms related to training perception, training effectiveness, and training transfer in organizational and software engineering contexts. The search identified 20 sources — including empirical studies, validated instruments, and related literature — from which questionnaire items or constructs were extracted. The primary sources included studies on training transfer and perception of learning \cite{facteau1995influence, santos2003employee, coverstone2003training, devaraj2004measure, fawad2012integrated, yaqoot2021predicted, baldwin1991perils}. Each extracted item was mapped to the corresponding subcategory of Salas' framework (Table~\ref{tab:salas_categories}), resulting in a matrix of 106 candidate items distributed across the four dimensions: Training Needs Analysis, Antecedent Conditions, Training Methods/Strategies, and Post-Training Conditions.

\textbf{Stage~2 — Semantic grouping (40 items).} Items addressing the same underlying construct were grouped, and redundancies across sources were eliminated. For instance, multiple items measuring ``perceived supervisor support for training'' from different studies were consolidated into a single representative item. This stage reduced the pool from 106 to 40 items while preserving coverage across all framework dimensions.

\textbf{Stage~3 — Likert scale adaptation (30 items).} The 40 grouped items were reformulated as declarative statements suitable for a five-point Likert scale (1=\textit{Strongly Disagree}; 5=\textit{Strongly Agree}), anchored to the participant's most recent learning experience. Open-ended and categorical questions were converted to perception-based statements. This reformulation reduced the set to 30 items.

\textbf{Stage~4 — Cognitive load reduction (27 items).} A cognitive load reduction process was applied to simplify wording, eliminate ambiguities, and shorten item length without altering the intended construct. Three items were removed for being redundant with other items after simplification. The resulting instrument comprised 27 closed Likert-scale items and 9 sociodemographic questions.

Table~\ref{tab:instrument_example} illustrates the refinement process for the construct \textit{Instructor Performance} (Salas' dimension: Training Methods and Instructional Strategies), which emerged as one of the strongest predictors of perceived quality (Q18, $\rho=0.785$).

\begin{table*}[ht]
    \centering
    \caption{Instrument refinement example for the construct \textit{Instructor Performance}.\protect\footnotemark}
    \label{tab:instrument_example}
    \small
    \renewcommand{\arraystretch}{1.4}
    \begin{tabular}{| p{0.14\linewidth} | p{0.80\linewidth} |}
        \hline
        \textbf{Stage} & \textbf{Item content} \\
        \hline
        \textbf{1. Extraction} \newline (106 items) &
        Multiple items from different sources: \textit{``Trainer was helpful''}, \textit{``Trainer was well prepared''}, \textit{``Training showed encouragement and motivated trainees to learn''}, \textit{``Trainer used varied learning methods''} \cite{fawad2012integrated}; \textit{``The trainer keeps current and up to date on the subject''} \cite{yaqoot2021predicted}; \textit{``Were the instructor's skills good?''} \cite{devaraj2004measure}. \\
        \hline
        \textbf{2. Grouping} \newline (40 items) &
        Consolidated into a single composite item: \textit{``The training instructor contributed significantly to my learning by using clear and varied methods, encouraging active participation, providing support, and facilitating practical application of the content.''} \\
        \hline
        \textbf{3. Adaptation} \newline (30 items) &
        Split into two focused Likert statements: (a)~\textit{``The instructor's explanation was clear and easy to understand.''}; (b)~\textit{``The instructor's performance was essential to my learning and motivation.''} \\
        \hline
        \textbf{4. Reduction} \newline (27 items) &
        Final items after cognitive load simplification: \textbf{Q17}~—~\textit{``The instructor's explanation was clear and easy to understand.''}; \textbf{Q18}~—~\textit{``The instructor's performance was essential to my learning and motivation.''} \\
        \hline
    \end{tabular}
\end{table*}

\footnotetext{Items in stages 2--4 are English translations; the original instrument was administered in Portuguese.}

Table~\ref{tab:instrument_items} presents the complete set of 27 Likert-scale items with their short labels and full statement wordings, organized by framework dimension.

\begin{table*}[ht]
    \centering
    \caption{Survey instrument: 27 Likert-scale items organized by Salas' framework dimensions.\protect\footnotemark}
    \label{tab:instrument_items}
    \small
    \renewcommand{\arraystretch}{1.25}
    \begin{tabular}{| c | p{0.18\textwidth} | p{0.68\textwidth} |}
        \hline
        \textbf{ID} & \textbf{Short label} & \textbf{Full statement} \\
        \hline
        \multicolumn{3}{|l|}{\textit{Training Needs Analysis}} \\
        \hline
        Q1  & Alignment with objectives     & I understood how the learning experience connected with the company's objectives. \\
        Q2  & Support and resources          & I had the necessary support and resources (time, technology, support) from the company for this learning experience. \\
        Q3  & Content aligned with job needs & The content of the learning experience met the needs of my role. \\
        Q4  & Adapted to experience level    & The content of the learning experience was adapted to my level of experience. \\
        Q5  & Consideration of employee opinion & My opinion was considered in designing the learning experience I attended. \\
        \hline
        \multicolumn{3}{|l|}{\textit{Antecedent Training Conditions}} \\
        \hline
        Q6  & Impact on personal time        & The workload of this learning experience affected my personal time (rest, social life). \\
        Q7  & Relevance for competitiveness  & The learning experience was important to keep me competitive in the market. \\
        Q8  & Motivation to learn            & What I learned in the learning experience motivated me to seek new knowledge. \\
        Q9  & Leadership encouragement       & My area's leadership actively encouraged my participation in the learning experience. \\
        Q10 & Training during working hours  & I had enough time during working hours to participate in the learning experience and study. \\
        Q11 & Obligation over interest       & I participated in this learning experience more out of obligation than interest in the content. \\
        Q12 & Incentives                     & The company offered clear incentives for participation in this learning experience (e.g., bonuses, time off, cost reimbursement, performance evaluation points). \\
        Q13 & Recognition                    & The company usually recognizes employees who develop and complete training (e.g., through certificates, public praise, announcements). \\
        Q14 & Organization and structure     & The learning experience I attended was well organized and structured. \\
        Q15 & Useful materials               & The materials I received for the learning experience were useful. \\
        \hline
        \multicolumn{3}{|l|}{\textit{Training Methods and Instructional Strategies}} \\
        \hline
        Q16 & Adequate environment           & The learning experience environment (physical or virtual) was adequate for learning. \\
        Q17 & Instructor clarity             & The instructor's explanation was clear and easy to understand. \\
        Q18 & Instructor performance         & The instructor's performance was essential to my learning and motivation. \\
        Q19 & Varied activities              & The activities during the learning experience were interesting and varied. \\
        Q20 & Soft skills focus              & The learning experience included a relevant part dedicated to the development of soft skills (behavioral skills). \\
        \hline
        \multicolumn{3}{|l|}{\textit{Post-Training Conditions}} \\
        \hline
        Q21 & Overall satisfaction           & I was satisfied with the quality of the learning experience I attended. \\
        Q22 & Problem-solving reasoning      & The learning experience helped me develop my reasoning for solving problems. \\
        Q23 & Performance improvement        & Applying what I learned in the learning experience improved my performance. \\
        Q24 & Practical applicability        & I can apply in my work what I learned in the learning experience. \\
        Q25 & Autonomy at work               & What I learned in the learning experience gave me more autonomy at work. \\
        Q26 & Leadership support for transfer & My leadership's support was an incentive for me to apply what I learned. \\
        Q27 & Career growth opportunities    & The learning experience I completed opened growth opportunities in the company (promotion or salary increase). \\
        \hline
    \end{tabular}
\end{table*}

\footnotetext{All items use a 5-point Likert scale (1=Strongly Disagree; 5=Strongly Agree). Statements are English translations; the original instrument was administered in Portuguese. The complete instrument in Portuguese is available in the replication package (Section~\ref{sec:artefact}).}

A pilot study with 10 software engineering professionals evaluated clarity and understanding, resulting in minor adjustments, most notably replacing the term \textit{training} with \textit{learning experience}. The final instrument is available as supplementary material in the replication package (Section~\ref{sec:artefact}).

\subsection{Participants and Sampling}

The target population comprises professionals working in software engineering in Brazil, with higher education in Computing or related Information and Communication Technology (ICT) areas. Recruitment was by convenience, through professional social networks (LinkedIn), messaging app groups (WhatsApp and Telegram), and email lists. Participation was voluntary, with no financial incentives.

Inclusion criteria:
\begin{itemize}
    \item being 18 years of age or older;
    \item working professionally in the software engineering area;
    \item having participated, in the last 12 months, in at least one corporate training sponsored by the employing organization.
\end{itemize}

\subsection{Data Collection Procedure}

Data collection took place entirely online, from September 9 to November 2, 2025 ($\approx$8 weeks). Completion of the consent, sociodemographic, and training characterization sections was mandatory. No data imputation techniques were adopted. One response was excluded for not meeting the inclusion criteria, resulting in \textbf{282} valid responses.

\subsection{Data Analysis}

Since the characterization items are ordinal, polychoric correlation was adopted, the recommended approach for Likert-type scales when estimating associations between latent variables underlying ordinal responses \cite{lorenzo2006factor, holgado2010polychoric}. Item \textbf{Q21} (Overall satisfaction) was treated as the reference variable for perceived training effectiveness and quality, and its correlations with all other items were examined to identify the principal predictors of perceived quality and effectiveness.

\subsection{Ethical Considerations}

The study was approved by the Research Ethics Committee (CAEE: 91121125.4.0000.5208; Opinion No.\ 7.816.810). Participation was voluntary, and data were collected anonymously and confidentially. Participants could withdraw at any time without prejudice.

\section{Results}\label{section:results}

\subsection{Sociodemographic and Professional Profile}
\label{sec:qs}

The valid sample ($n=282$) is characterized as follows.

Regarding \textbf{gender}, 76.6\% ($n=216$) identified as male, 23.0\% ($n=65$) as female, and one participant (0.4\%) chose not to respond.

The \textbf{age distribution} is unimodal and approximately symmetric ($mean=33.78$, $SD=7.73$, $median=33$ years).

Regarding \textbf{educational level}, 49.7\% ($n=140$) hold a completed undergraduate degree and 38.7\% ($n=109$) have completed postgraduate education; only 11.7\% ($n=33$) have not yet completed higher education.

Regarding \textbf{organization size}, 77.3\% ($n=218$) work in companies with 100 or more employees; medium-sized companies (50--99 employees) account for 11.7\% ($n=33$) and small companies (10--49 employees) for 8.5\% ($n=24$). No professionals from micro-enterprises participated.

In terms of \textbf{area of activity}, Development dominates (44.0\%, $n=124$), followed by People/Project Management (14.5\%, $n=41$), Quality Assurance (12.4\%, $n=35$), and Technical Leadership (8.5\%, $n=24$). Data Science (6.4\%), DevOps (6.0\%), and Architecture (3.9\%) are also represented.

Regarding \textbf{professional experience}, 43.6\% ($n=123$) have more than eight years in Software Engineering. The Senior category is the most frequent professional level (28.7\%, $n=81$), followed by Mid-level (19.9\%, $n=56$) and Leadership/Management (19.5\%, $n=55$). Entry-level positions account for 16.4\% ($n=46$).

Finally, most participants (73\%, $n=206$) reported that participation in the most recent training was by personal choice, taking advantage of a company-sponsored benefit.

\subsection{Training Characterization: Likert-Scale Responses}
\label{subsection:training}

\begin{figure*}[h]
    \centering
    \includegraphics[width=\textwidth]{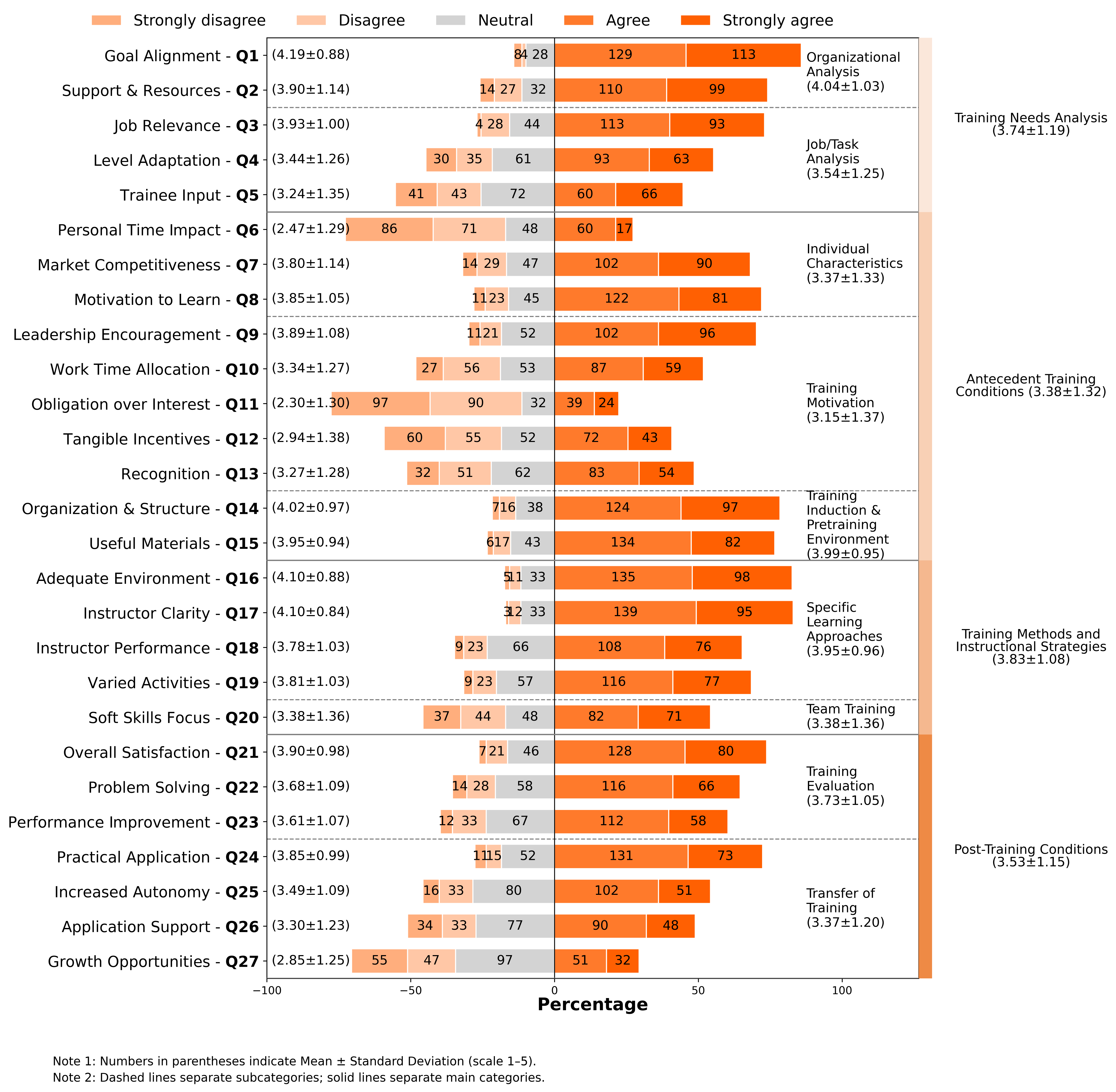}
    \caption{Distribution of Likert-scale responses by item and framework dimension.\protect\footnotemark}
    \label{fig:likert}
\end{figure*}

Participants evaluated their most recent learning experience using a five-point Likert scale, structured according to the four dimensions of \citet{salas2001science}'s framework: (i)~Training Needs Analysis; (ii)~Antecedent Conditions; (iii)~Training Methods and Instructional Strategies; and (iv)~Post-Training Conditions. The distribution of responses is illustrated in Figure~\ref{fig:likert}.

\footnotetext{Scale: 1=Strongly Disagree to 5=Strongly Agree. Items are shown as abbreviated labels; full wordings and framework dimensions are presented in Table~\ref{tab:instrument_items}.}

Overall, responses tend toward a positive evaluation, with mean values predominantly above the scale midpoint (3.0). Each item is identified by its short label as defined in Table~\ref{tab:instrument_items}. Notable item-level findings are as follows:

\textbf{Q1} (Alignment with objectives): High mean (4.188), indicating near-unanimous positive perception.

\textbf{Q5} (Consideration of employee opinion): Mean near the midpoint (3.238), reflecting substantial variation in feedback-incorporation practices across organizations.

\textbf{Q6} (Impact on personal time): Mean of 2.472, indicating that most participants perceive training as affecting their personal time, regardless of training quality.

\textbf{Q10} (Training during working hours): Mean near the midpoint (3.337), revealing heterogeneity in organizational policies for allocating work hours to learning.

\textbf{Q11} (Obligation over Interest): Low mean (2.301) with a right-skewed distribution, consistent with the demographic finding that 73.1\% of participants chose training voluntarily (QS1).

\textbf{Q12} (Incentives): Mean slightly below the midpoint (2.940), indicating divided perceptions of how organizations incentivize training participation.

\subsection{Correlation Analysis}
\label{sec:corr}

\begin{figure*}[h]
    \centering
    \includegraphics[width=\textwidth]{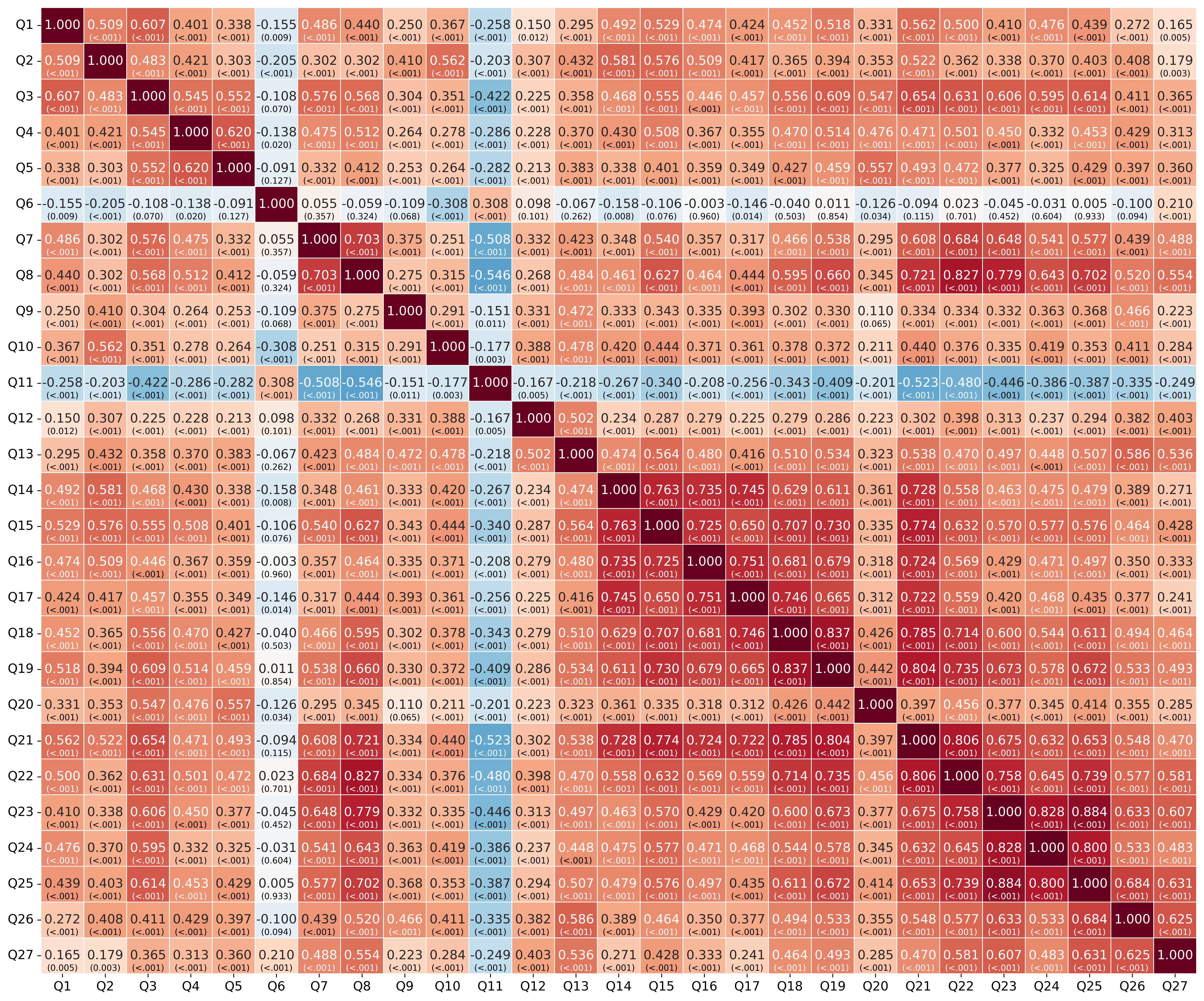}
    \caption{Polychoric correlation matrix ($N=282$).\protect\footnotemark}
    \label{fig:correlacao}
\end{figure*}

\footnotetext{Upper values: polychoric correlation coefficient ($\rho$); values in parentheses: $p$-value.}

The polychoric correlation matrix (Figure~\ref{fig:correlacao}) reveals three key patterns.

\textbf{Q6 (Impact on personal time)} shows a divergent pattern, with coefficients below the 0.30 threshold recommended by \cite{hair2009analise} for most pairs. Its only notable correlation was with \textbf{Q11} (Obligation over Interest; $\rho=0.308$, $p<0.001$), \textbf{\textit{suggesting, though weakly, that the perceived burden on personal time increases under mandatory participation.}}

\textbf{Q11 (Obligation over Interest)} showed a divergent pattern, with negative correlations with most variables. The strongest inverse associations were with \textbf{Q8} (Motivation to learn; $\rho=-0.546$, $p<0.001$), \textbf{Q21} (Overall satisfaction; $\rho=-0.523$, $p<0.001$), and \textbf{Q7} (Relevance for competitiveness; $\rho=-0.508$, $p<0.001$). \textbf{\textit{These results provide statistical evidence that perceived mandatory participation reduces both motivation and perceived training quality.}}

\textbf{Q21 (Overall satisfaction)} emerged as the central indicator of perceived effectiveness. Its strongest associations were with \textbf{Q22} (Problem-solving reasoning; $\rho=0.806$, $p<0.001$), \textbf{Q19} (Varied activities; $\rho=0.804$, $p<0.001$), and \textbf{Q18} (Instructor performance; $\rho=0.785$, $p<0.001$). \textbf{\textit{These findings show that cognitive impact, activity variety, and instructor performance are the central drivers of perceived training quality and effectiveness.}}

In contrast, the correlation between \textbf{Q21} and \textbf{Q6} was weak and non-significant ($\rho=-0.094$, $p=0.115$). \textbf{\textit{This suggests that personal time investment is largely independent of training quality.}}

Analysis scripts and the full pipeline for reproducing the polychoric correlation matrix are available in the public repository described in the Artifact Availability section (Section~\ref{sec:artefact}).

\section{Discussion}\label{section:discussion}

This section discusses the quantitative findings in light of the literature on organizational training. The results are interpreted within the context of software engineering, where rapid technological change \cite{assyne2022state} and a structural academia--industry skill gap \cite{diniz2024skill} shape corporate training dynamics. The discussion is organized by research question.

\subsection*{RQ1 — Factors Determining Training Quality Perceptions}

In response to RQ1, the strongest predictors of perceived training quality and effectiveness are problem-solving reasoning (Q22, $\rho=0.806$), variety of activities (Q19, $\rho=0.804$), and instructor performance (Q18, $\rho=0.785$). These three factors form a tightly connected core that accounts for the highest associations with perceived training quality. These findings are consistent with \citet{fawad2012integrated}, who position instructor performance and content relevance as key dimensions of training effectiveness, and with \citet{devaraj2004measure}, who highlight direct applicability to job performance as a central determinant of perceived training value. More broadly, \citet{salas2012science} identify active learning, practice opportunities, and feedback as evidence-based principles that enhance training outcomes. These principles align with the cognitive impact and activity variety factors observed in this study.

The strong role of instructor performance ($\rho=0.785$) aligns with \citet{yaqoot2021predicted}, who show that instructor competence is especially influential in face-to-face and technical training. This underlines the importance of investing not only in curriculum design but also in the pedagogical and technical preparation of instructors.

The emergence of these three factors as the strongest predictors of perceived training quality finds theoretical support in \citet{bandura1977social}'s observational learning model. Bandura identifies four subprocesses for effective learning: attention, retention, reproduction, and motivation. The instructor, as the primary model in training, governs the attention process through clarity and engagement (Q17, Q18). Problem-solving reasoning (Q22) reflects the retention process, where learners encode and internalize observed knowledge through cognitive engagement. Varied activities (Q19) sustain motivation by providing diverse stimuli and reinforcement throughout the learning experience. This alignment suggests that the three-factor core identified in this study is not merely a statistical pattern, but reflects fundamental mechanisms of human learning as described by social learning theory.

The Job/Task Analysis subcategory, identified as a gap by \citet{de2025mapping} and positioned by \citet{salas2001science} as the essential foundation for effective training, showed near-neutral results overall (mean=3.54), yet the item on alignment with job function needs scored high (mean=3.93). This suggests that participants value content connected to their daily tasks, even when formal needs analysis processes are absent. Items on adapting training to experience level (mean=3.44) and considering professional opinions in training design (mean=3.24) indicate clear room for improvement.

Finally, Q6 (personal time burden) showed no significant correlation with Q21 (overall satisfaction), suggesting that reducing time cost alone is unlikely to increase perceived training quality without improving instructional quality.

\subsection*{RQ1.1 — Effect of Mandatory Versus Voluntary Participation}

In response to RQ1.1, the results indicate that mandatory participation negatively influences motivation, perceived relevance, and perceived training quality, while also amplifying the perception of personal time burden.

The correlation data clearly support the negative effect of mandatory participation on both motivation and perceived training quality. The strong inverse correlations between Q11 (Obligation over Interest) and motivation (Q8), perceived quality (Q21), and perceived relevance (Q7) align with \citet{gegenfurtner2016voluntary}, who show that autonomous motivation, linked to voluntary participation, leads to better training reactions and transfer. This is also consistent with \citet{salas2012science}, who identify pretraining motivation as one of the most critical antecedents of learning outcomes, noting that organizational factors, including how participation is framed, directly shape this motivation. The additional correlation between Q6 and Q11 suggests that mandatory training also increases the perceived burden on personal time.

However, the literature cautions against fully endorsing voluntary training. \citet{baldwin1991perils} and \citet{tsai2003perceived} argue that too much voluntariness can signal low organizational commitment to training, weakening its perceived strategic value. The implication is that organizations should aim for balance: positioning training as strategically relevant while preserving as much participant autonomy as possible.

\subsection*{RQ1.2 — Consistency with the General Training Literature}

In response to RQ1.2, the results show a notable consistency with the general training literature. The three factors most strongly correlated with perceived training quality align with established training effectiveness research \cite{salas2001science, salas2012science, fawad2012integrated, yaqoot2021predicted, devaraj2004measure}. Similarly, the negative effect of mandatory participation on motivation and perceived quality converges with evidence from non-SE populations \cite{gegenfurtner2016voluntary, de2025effects, curado2015voluntary}, and the role of social support from supervisors and peers aligns with \citet{facteau1995influence} and \citet{santos2003employee}.

This consistency suggests that established training design guidelines (investing in instructor qualification, designing engaging activities, and preserving participant autonomy) are applicable to SE professionals as they are to other knowledge workers. This pattern mirrors the experience in healthcare, where \citet{salas2009critical} found that the same organizational factors (leadership support, instructor qualification, trainee motivation, and learning climate) proved critical for training success despite the domain-specific characteristics of medical teams. What remains open for future research is whether the \textit{relative weights} of these factors differ across domains, for example whether problem-solving reasoning carries even more weight in SE given the technical nature of the work, or whether the mandatory--voluntary dynamic works differently for hard-skill versus soft-skill training, as suggested by \citet{de2025effects}. Cross-domain comparative studies, such as between SE and healthcare professionals using the same instrument, would provide stronger evidence on this question.

\subsection*{Salas' Framework as an Exploratory Lens for SE}

Salas and Cannon-Bowers' framework \cite{salas2001science}, originally from organizational psychology, produced coherent and interpretable results when applied to the SE context. The framework has already been successfully applied in healthcare \cite{salas2009critical}, and since \citet{de2025mapping} found no SE-specific instruments for assessing training effectiveness, this operationalization was a necessary first step. The instrument produced interpretable distributions across all four framework dimensions, and the main findings align with the broader training literature \cite{salas2001science, salas2012science, fawad2012integrated, gegenfurtner2016voluntary, yaqoot2021predicted}. However, this study did not conduct reliability analysis (Cronbach's alpha or McDonald's omega) or exploratory factor analysis, so no claims about construct validity or internal consistency can be made at this stage. The subcategories \textit{Simulation-Based Training and Games} and \textit{Team Training} showed limited variance, suggesting these dimensions may need adaptation for SE training practices. The instrument should therefore be treated as an exploratory candidate requiring future psychometric validation, including reliability analysis, exploratory and confirmatory factor analysis, and cross-cultural replications.

\section{Threats to Validity and Limitations}
\label{sec:limitations}

\textbf{Convenience sampling}: Recruitment via professional networks and messaging groups may introduce self-selection bias, potentially over-representing professionals who are already engaged with training.

\textbf{Sample restricted to Brazil}: Results may not generalize to other cultural or economic contexts.

\textbf{Memory bias}: Participants reported their most recent training experience, which may be subject to recency effects or forgetting.

\textbf{High educational profile}: 88.3\% hold a complete undergraduate or postgraduate degree, which may not reflect the full Brazilian software workforce.

\textbf{Gender imbalance}: The sample is 76.6\% male, which limits the power of gender-disaggregated analyses. Future work should use stratified sampling to achieve more balanced representation.

\textbf{Measurement validity}: Professional level categories (e.g., Junior/Senior) may be interpreted heterogeneously. Future replications should adopt explicit operational definitions for each level.

\textbf{Construct validity of single-item measures}: Each construct (e.g., perceived quality, instructor performance) was assessed with a single Likert-scale item, which may not fully capture its multidimensional nature. While single-item measures are common in large-scale surveys, they limit internal consistency assessment. Future studies should use multi-item scales and confirmatory factor analysis.

\textbf{Absence of objective metrics}: This study relies on self-reported perceptions. Future work should complement survey data with objective effectiveness measures such as productivity or code quality indicators.

\textbf{Role imbalance and absence of subgroup analysis}: Developers make up 44\% of the sample ($n=124$), which may bias findings toward technical training perspectives. This study does not test whether perceptions differ across professional roles, organization sizes, or other sociodemographic variables. A companion study currently under review addresses this gap by testing 243 combinations of perception items and sociodemographic variables on the same dataset, finding that training effectiveness depends more on instructional design than on participant profile.

\textbf{Insufficient differentiation of training types}: The survey does not distinguish between different types of corporate training, such as onboarding, upskilling courses, team-building workshops, or compliance sessions. These formats differ in objectives, duration, and pedagogy, which likely influences perceptions. Because respondents evaluated their ``most recent learning experience'' without further characterization, findings may mix different training experiences. Future studies should add typological controls or stratified analyses.

\textbf{Use of Salas' framework}: The framework was originally developed in organizational psychology and has not been psychometrically validated for the SE context. No reliability analysis (Cronbach's alpha or McDonald's omega) or factor analysis was conducted on the data. The instrument produced in this study is exploratory and should be treated as a candidate for future psychometric validation, not as a validated measurement tool.

\textbf{Data confidentiality}: Individual-level response data cannot be shared publicly due to the confidentiality agreement with the Research Ethics Committee (CAEE: 91121125.4.0000.5208). The questionnaire, analysis scripts, and aggregated results are available as open-access artifacts (Section~\ref{sec:artefact}). Researchers interested in accessing the raw dataset should contact the corresponding author.

\section{Conclusion}\label{section:conclusion}

This study quantitatively analyzed the perceptions of 282 Brazilian software engineering professionals regarding corporate training quality and effectiveness, operationalizing Salas and Cannon-Bowers' framework through a large-scale Likert-scale survey and polychoric correlation analysis. The main findings are summarized as follows.

First, regarding \textbf{RQ1}, three tightly correlated factors emerged as the strongest predictors of perceived training quality and effectiveness: cognitive engagement (Q22, $\rho=0.806$), variety of activities (Q19, $\rho=0.804$), and instructor performance (Q18, $\rho=0.785$). When resources are limited, organizations should prioritize these three dimensions. Notably, perceived personal time burden showed no significant correlation with training quality, suggesting that improving scheduling alone is unlikely to enhance perceived quality without gains in instructional design.

Second, regarding \textbf{RQ1.1}, mandatory participation negatively influences motivation, perceived relevance, and perceived training quality, while also amplifying the perception of personal time burden. Organizations should preserve participant autonomy and clearly communicate the value of training, as mandatory participation consistently weakens engagement.

Third, regarding \textbf{RQ1.2}, these findings are consistent with the general training literature, suggesting that software engineering organizations do not need to reinvent training design principles. Established guidelines from other domains can be applied to the software industry with confidence.

Fourth, the operationalization of Salas and Cannon-Bowers' framework produced coherent and interpretable results in the SE context, suggesting it is a good candidate as an analytical lens for this domain. However, since no reliability or factor analysis was conducted, the instrument remains exploratory and requires future psychometric validation, including reliability analysis (Cronbach's alpha or McDonald's omega), exploratory and confirmatory factor analysis, and cross-cultural replications.

A companion study, currently under review, examines whether these patterns hold across different professional roles, experience levels, and organizational contexts by applying significance tests across sociodemographic variables on the same dataset.

\section*{Declaration of the Use of Artificial Intelligence}
This research was originally developed in Portuguese. The authors used Generative AI tools (OpenAI ChatGPT, Google Gemini, and Anthropic Claude) exclusively for support in translating to English, improving textual cohesion and clarity, and assisting with the structuring and revision of the manuscript.

\section*{Artifact Availability}\label{sec:artefact}
The questionnaire, the question design matrix documenting all four instrument development stages, analysis scripts, and results visualizations are available as open-access artifacts at \href{https://doi.org/10.5281/zenodo.19172188}{Zenodo (10.5281/zenodo.19172188)}.

\bibliographystyle{ACM-Reference-Format}
\bibliography{qualificacao}

\end{document}